\documentclass{PoS}

\usepackage{amsfonts, amssymb, amsmath, graphicx, color}

\newcommand{\beq}{\begin{equation}}
\newcommand{\eeq}{\end{equation}}
\newcommand{\calH}{ {\cal H} }

\newcommand{\rmi}{ {\rm i} }

\title{Delineating the properties of neutron star matter in cold, dense QCD}
	\ShortTitle{Delineating the properties of matter in cold, dense QCD}
\author{\speaker{Toru Kojo}\\%
      Central China Normal University\\
      E-mail: \email{torujj@mail.ccnu.edu.cn}}
\abstract{
The properties of dense QCD matter are delineated through the construction of equations of state which should be consistent with the low and high density limits of QCD, nuclear laboratory experiments, and the neutron star observations. These constraints, together with the causality condition of the sound velocity, are used to develop the picture of hadron-quark continuity in which hadronic matter continuously transforms into quark matter (modulo small 1st order phase transitions). The resultant unified equation of state at zero temperature and $\beta$-equilibrium, which we call Quark-Hadron-Crossover (QHC19), is consistent with the measured properties of neutron stars as well as the microphysics known for the hadron phenomenology. In particular to $\sim 10n_0$ ($n_0$: saturation density) the gluons remain as non-perturbative as in vacuum and the strangeness can be as abundant as up- and down-quarks at the core of two-solar mass neutron stars. Within our modeling the maximum mass is found less than $\simeq 2.35$ times solar mass and the baryon density at the core ranges in $\sim 5$-8$n_0$.

}
\FullConference{37th International Symposium on Lattice Field Theory - Lattice2019 \\
16-22 June 2019\\
Wuhan, China}

\begin{document}	

%%%%%%%%%%%%%%%%%
\section{Introduction}
%%%%%%%%%%%%%%%%%

%%%%%%%%%
\begin{figure}
\begin{center}
\vspace{-1.0cm}
\hspace{-2.5cm}
\includegraphics[width=6.50cm,bb=0 0 600 600, angle =0]{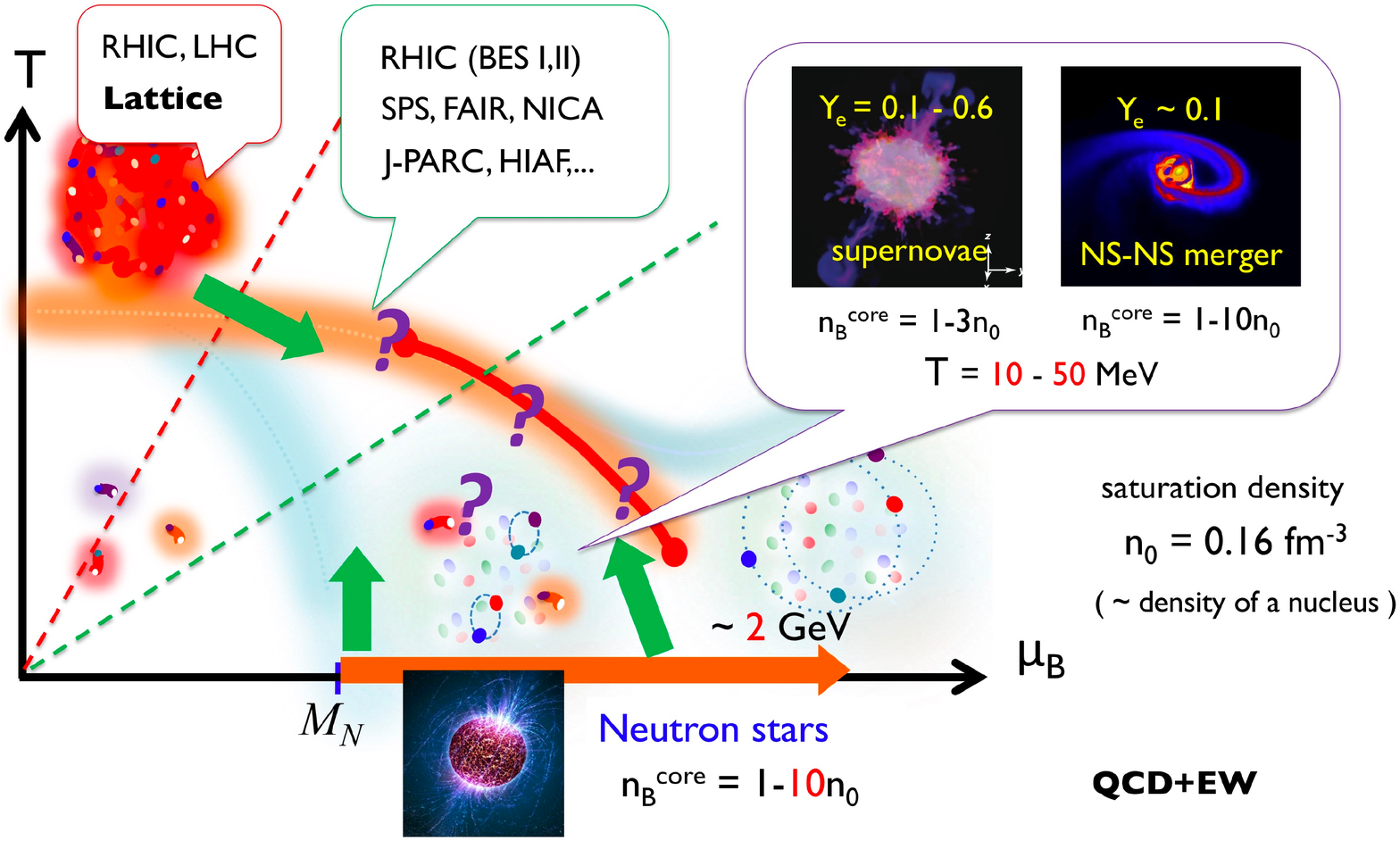}
\vspace{-0.3cm}
\caption{
\footnotesize{ The phase diagram for QCD (+EW) in the $\mu_B$-$T$ plane. The physics of neutron stars is relevant for high baryon density to $\sim 10n_0$ and temperature from keV to $\sim 50$-$100$ MeV. The chemical potentials for electric charge and strangeness are not shown but very important in neutron stars.
}
  }
  \vspace{-0.5cm}
  \end{center}
 \label{fig:QCD_phase_diagrram}
\end{figure}
%%%%%%%%%

I have been asked by the organizers of this conference to give a brief summary of recent developments in the physics of neutron stars (NSs) and to phrase those findings in the QCD context. 
The goal of this talk is to identify the relevant microphysics and effective interactions which may have strong connection to the quark dynamics at semi-short distance, e.g., the dynamics inside of hadrons as well as the baryon-baryon interactions at short distance. The lattice QCD simulations can offer these microscopic information which can be assembled to constrain our descriptions for, e.g., the QCD equations of state (EoS).

Discussions of NSs have long history but last 10 years have witnessed dramatic progress driven mainly by the discoveries of two-solar mass ($2M_\odot$) NS \cite{fonseca,Demorest,Antoniadis2013,Cromartie:2019kug} and the historical detections of gravitational waves (GWs) from the NS-NS merger GW170817 \cite{TheLIGOScientific:2017qsa,GW170817A} (and more recently hints of NS-BH mergers \cite{Lattimer:2019qdc}).
These findings have changed the conditions required for EoS, and a number of new EoS have been constructed to satisfy the new observational constraints. Such construction is based on a variety of assumptions on the effective degrees of freedom as well as their interactions. In particular the degrees of freedom used for the high density part of the EoS can be very different; some include quark matter in the core, some do not, but they can give similar neutron star structures by adjusting interactions. Accordingly the distinction of NS with and without quark matter core cannot be achieved by just looking at the neutron star structure; the discussion based on the microphysics is necessary to disentangle the degeneracy. For this reason this talk was dedicated for the microphysics at baryon density ($n_B$) relevant for the neutron star physics. We take the saturation density $n_0 \simeq 0.16\, {\rm fm}^{-3}$ as our unit for $n_B$.

In the context of the QCD phase diagram (Fig.\ref{fig:QCD_phase_diagrram}), NS are unique cosmic laboratories to test our understanding of QCD matter at high baryon density. The core density of static NS is $n_B = 1$-$10n_0$ (or $\mu_B =1$-$2$ GeV) and the temperature $T$ is the order of keV. For the dynamical phenomena such as supernovae or NS mergers, the temperature may reach $T\sim 50$ MeV at low density and $T\sim 20$ MeV at high density, unless the significant latent heat is produced via hypothetical first order phase transitions \cite{Most:2018eaw}. These domains are not directly accessible by the current lattice QCD simulations because of the sign problem. Meanwhile the heavy ion collisions at low energy can create highly compressed matter reaching $\sim 5n_0$, 
but it also produces heat with which the resulting temperature is $\sim 50$-$100$ MeV, leaving small overlap with NS (isospin and strangeness chemical potentials are also different) except low density part of $\sim 1$-$2n_0$ of supernovae or NS mergers \cite{Motornenko:2019lwh}. 
In this respect NS is regarded as independent sources of constraints to understand the global structure of the phase diagram at high density from the low temperature side.

This talk was structured as follows. We begin with some theoretical orientations and then review the NS constraints with the emphasis on their implications for the QCD EoS. This supposed EoS will be used to delineate the properties of matter in NS domains. The relevant interactions in the quark dynamics are discussed by quoting quark descriptions of hadron physics and baryon-baryon interactions, which can be explored by the lattice simulations.

This talk is mainly based on our reviews \cite{Baym:2017whm,Kojo:2015fua} and the update \cite{Baym:2019iky,Kojo:2019raj} since then. The tables of our EoS, QHC19, are available on the Web at [http://www.np.phys.waseda.ac.jp/EOS/] and the CompOSE archive at [https://compose.obspm.fr/eos].

%%%%%%%%%%%%%%%%%
\section{Theoretical orientations}
%%%%%%%%%%%%%%%%%

The high density limit of QCD EoS should be described well with weak coupling methods. The EoS is a bulk quantity which is sensitive to typical interactions, in this case large momentum transfer of the quark Fermi momentum, $p_F \sim \mu_B/3$. The pQCD calculations for quark matter EoS has been completed to 3-loop order  \cite{Kurkela:2009gj,Fraga:2001id,Freedman:1976ub}, and partial resummation of even higher order graphs \cite{Gorda:2018gpy}. The examination of $\alpha_s$ expansion or renormalization scale dependence tell us that the results become sensitive to the infrared contributions at quark chemical potential $\mu_q = \mu_B/3 \lesssim 1$ GeV, or in terms of density $\lesssim 50$-$100\, n_0$. Thus quark matter at $\lesssim 50\, n_0$ should be strongly correlated.

Further hint for quark matter comes from the dense regime of 2-color QCD for which the lattice simulations can be performed without suffering from the sign problem. The phase diagram, EoS, diquark condensates, Polyakov loops, and so on, in medium have been computed  \cite{Hands:2010gd,Cotter:2012mb,Braguta:2016cpw}. There are also results for the in-medium gluon propagators \cite{Hajizadeh:2017ewa,Boz:2018crd}. This theory has three distinct phases, hadronic, quark-gluon-plasma, and quark matter phases with (color-singlet) diquark condensates. With the scale setting by the string tension to the pure Yang-Mills case, the transition temperature for diquark condensates is about $T_c \sim 80$-$120$ MeV and is not very sensitive to baryon density to $\mu_q \sim 1$ GeV. As the $T_c$ is usually the order of the diquark gap, the lattice results indicate that the nonperturbative effects remain substantial at such large density. Recently the nonperturbative gluon propagators in medium are studied in analytic methods \cite{Suenaga:2019jjv} and compared to the lattice results \cite{Boz:2018crd}. The study indicates that the gluon propagators remain nonperturbative to $\mu_q \sim 1$ GeV.

Now we turn to the low density. In the dilute regime the nuclear matter description should work. The microscopic many-body calculations based on the empirically determined 2-body and 3-body interactions describe the nuclear matter properties at $n_B \simeq n_0$ well within the uncertainty in short range 3-body forces \cite{Akmal:1998cf,Togashi:2017mjp}. In slightly denser regime beyond $n_B\simeq 2n_0$, however, there arise questions in the systematics; the 3-body contributions grow faster than the 2-body ones. In fact their sizes are comparable already at $n_B \simeq 2n_0$, and at higher density 3- and even more-body forces dominate the EoS. Within pure nuclear modeling, the growth of 3-body contributions typically results in the speed of sound $c_s=( \partial P/\partial \varepsilon)^{1/2}$ ($P$: pressure; $\varepsilon$: energy density) greater than the speed of light somewhere around $\sim 5$-$10n_0$ \cite{Akmal:1998cf,Togashi:2017mjp}. Introduction of 4- or 5-body forces would temper such unphysical growth, but then we also {\it must} expect the importance of even more-body forces. To find fundamental resolutions we need to look into the origin of many-body forces based on the quark dynamics. 

%%%%%%%%%
\begin{figure}
\begin{center}
\vspace{-1.0cm}
\hspace{-2.5cm}
\includegraphics[width=7.0cm,bb=0 0 600 600, angle =0]{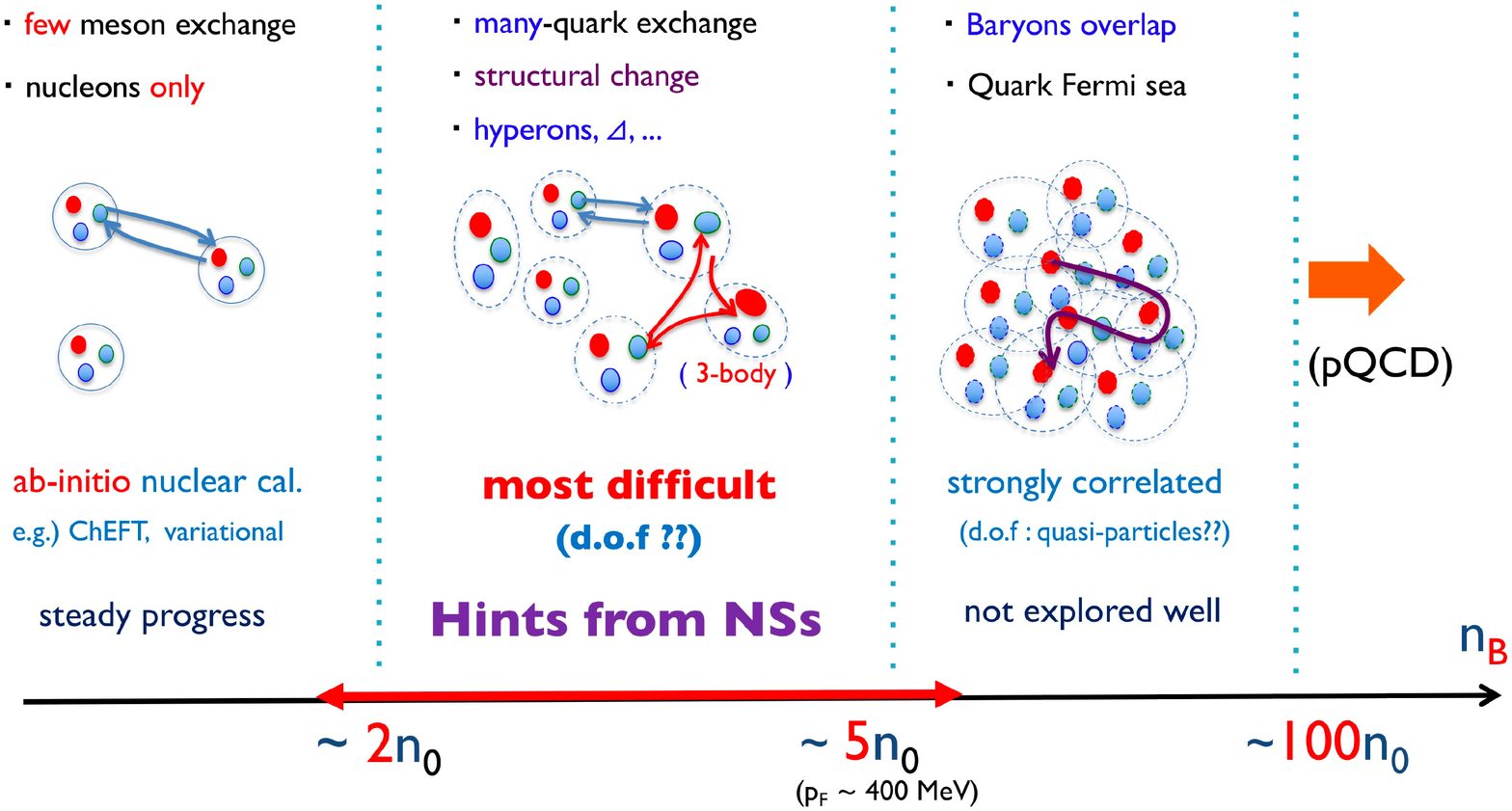}
\vspace{-1.cm}
\caption{
\footnotesize{ A 3-window description of QCD matter.
}
  }
  \vspace{-0.5cm}
  \end{center}
 \label{fig:3-window}
\end{figure}
%%%%%%%%%

With these findings we apply the following picture by decomposing the domain of NS matter into 3-windows \cite{Kojo:2015fua,Masuda:2012kf,Masuda:2012ed,Kojo:2014rca}; 
the nuclear regime at $n_B \le 2n_0$; the quark regime at $n_B \ge 5n_0$; and the intermediate regime for $2$-$5n_0$. The schematic picture is illustrated in Fig.\ref{fig:3-window}. In the dilute regime at low density, $n_B \le 2n_0$, baryons remain well-defined objects and the EoS are described by nuclear ones. From $\sim 2n_0$ to $\sim 5n_0$, it is unlikely that nucleons remain effective degrees of freedom but the density is not high enough for quarks to get fully released from baryons. At $n_B \sim 5n_0$, baryons with the radii of $\sim 0.5$ fm start to touch one another, and quarks should travel among different baryons. The formation of the quark Fermi sea should take place here. In three-flavor matter the quark Fermi momentum is $p_F \sim 400$ MeV (for two-flavor matter $p_F$ is even larger).

Theoretically the most uncertain is the domain of $n_B=2$-$5n_0$. The difficulty is largely due to the confining effects which transform the effective degrees of freedom. Meanwhile this domain is most important in the physics of NS. Fortunately the observations in the NS physics have improved significantly and now we have strong constraints on the properties of strongly correlated matter. For this reason we briefly review the constraints from recent NS observations.

%%%%%%%%%%%%%%%%%
\section{The neutron star constraints}
%%%%%%%%%%%%%%%%%

The fundamental quantity of NS is the mass-radius ($M$-$R$) relation. To calculate it we solve the Einstein equation coupled to QCD (+electroweak) EoS. The complexity depends on the assumptions. In most cases we may treat NS as static and spherical objects, and the Einstein equation is reduced to the Tolman-Oppenheimer-Volkoff (TOV) equation. EoS enters the TOV equation in the form of $P(\varepsilon)$. The $M$-$R$ relation and the EoS have one-to-one correspondence \cite{Lindblom1992} and in principle one can determine the QCD EoS directly from observed $M$-$R$ relations. In reality there are errors in measured $M$-$R$ relations, and the inversion procedures with weighted probability are necessary, see for instance \cite{Fujimoto:2017cdo,Fujimoto:2019hxv}.

If the rotational frequency of NS is large, we need to go beyond the TOV equation. If the frequency is not very large, the star rotates like a rigid body. The resulting maximum mass increases by $\sim 20\%$ by the rotational effects. This uniform rotation regime becomes invalid at larger frequency with which the rotation becomes differential. The latter is important when we discuss NS mergers. The maximum mass increases by $\sim 50\%$.

In order to characterize the structure of NSs we usually discuss the stiffness of EoS. In this talk ``stiff" EoS refers to that with large pressure $P$ at given energy density $\varepsilon$. The stiffer EoS generally lead to larger maximum masses and larger radii for NS; the energy density attracts the material to the center while the pressure increased in the compressed matter prevents the matter from falling. The speed of sound $c_s=(\partial P/\partial \varepsilon)^{1/2}$ is often used as the measure of the stiffness, but it is sometimes misleading as the derivative does not specify where $P(\varepsilon)$ starts. Indeed, if we start with a very stiff initial condition for $P(\varepsilon)$, even ideal gas EoS with $c_s^2=1/3$ can generate very large maximum masses as in stable pure quark star models \cite{Witten:1984rs}. 

To properly understand implications of the recent observations, it is crucial to specify which density domains are stiff, as we will discuss shortly. We will use the terminology ``soft-to-stiff", by which we mean that the EoS is soft at low density, $n_B \le 2n_0$, and stiff at high density, $n_B \ge 5n_0$. For the reasons described below, EoS leading to $R_{1.4} \le 13$ km for $1.4M_{\odot}$ stars will be called ``soft at low density", and EOS leading to $M \ge 2M_{\odot}$ will be called ``stiff at high density". Then the soft-to-stiff EoS generate the $M$-$R$ curves with the typical radii of $R_{1.4} \le 13$ km and the maximum mass $\ge 2M_\odot$. The terminology of the other combinations, such as ``stiff-to-stiff", ``stiff-to-soft", etc., should be already clear from this explanation.

%%%%%%%%%
\begin{figure}
\begin{center}
\vspace{-1.0cm}
\hspace{-2.5cm}
\includegraphics[width=7.0cm,bb=0 0 600 600, angle =0]{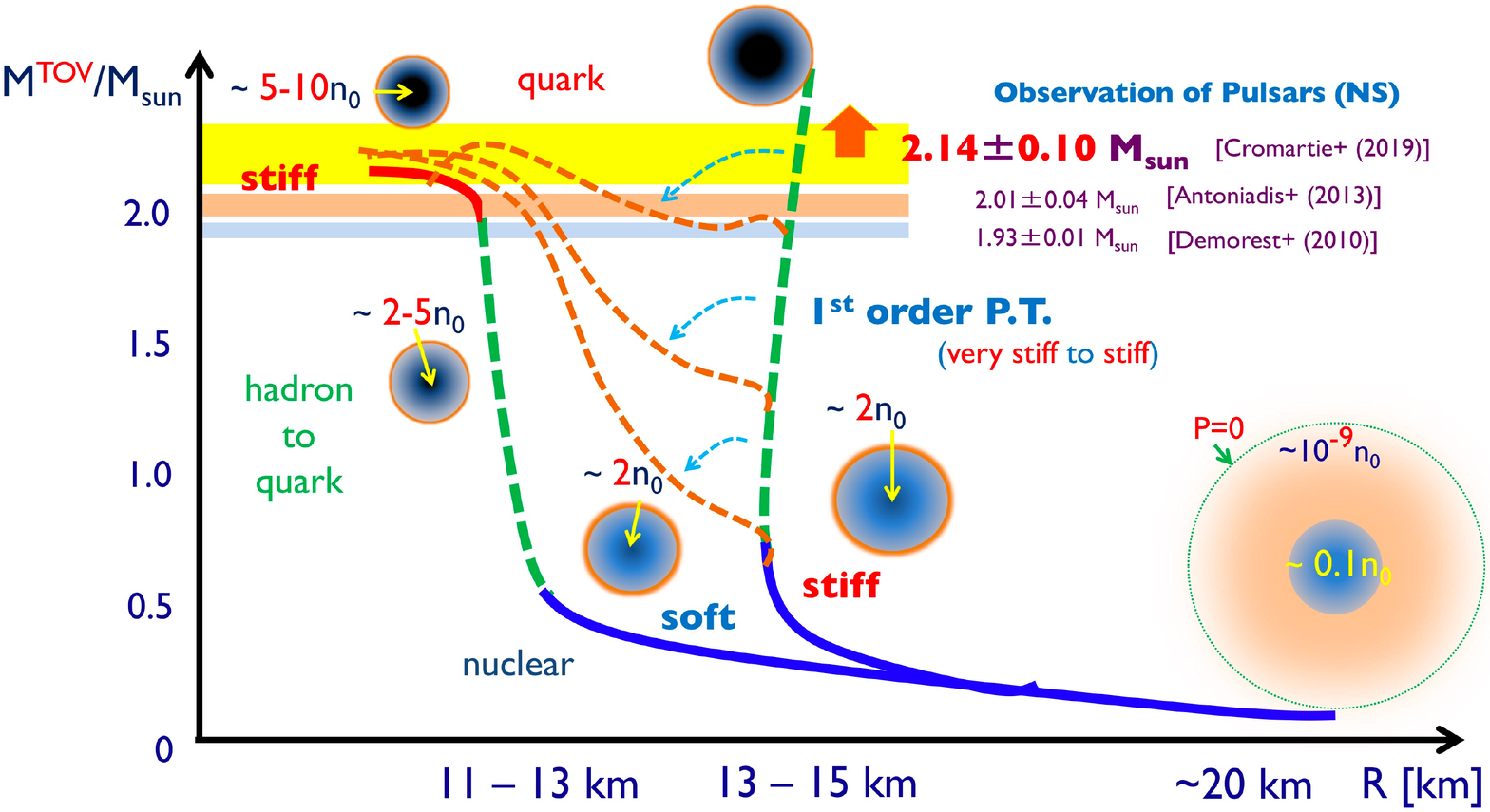}
\vspace{-1.cm}
\caption{
\footnotesize{ The $M$-$R$ relation for EoS.
}
  }
  \vspace{-0.5cm}
  \end{center}
 \label{fig:M-R}
\end{figure}
%%%%%%%%%

It has been known \cite{Lattimer:2006xb}  that the structure of $M$-$R$ curves has strong correlations with EoS at several fiducial densities, see Fig.\ref{fig:M-R}. For a low mass NS with the low density core, the material outside of the core is loosely bound by the gravity, so the corresponding radius, defined at the point of $P=0$, is large. For a slightly more massive NS, the radius is significantly smaller, because small increase of the gravity can compress the loosely bound material drastically. This trend continues until the loosely bound material becomes very thin, and then we observe the radius of the high density core dominated by a nuclear liquid at $n_B =1$-$2n_0$ with the repulsive forces. As a result the matter is no longer compressed substantially so that the $M$-$R$ curve goes vertically. Eventually the curve reaches the maximum mass $M_{\rm max}$. 

EoS at $n_B \gtrsim 5n_0$ must be sufficiently stiff; otherwise the star collapses to BH before reaching the observed lower bound $\simeq 2M_\odot$ for $M_{\rm max}$. The currently most stringent constraint is from the PSR J0740+6620 with $M = 2.14 \pm 0.10 M_\odot$ (at 68.3\% confidence level) \cite{Cromartie:2019kug} which have been reported in 2019. Before this announcement the bound was given by $1.928\pm0.017 M_\odot$ for the PSR J1614-2230 \cite{fonseca} and $2.01 \pm 0.04 M_\odot$ for the PSR J0348+0432 \cite{Antoniadis2013}. 
%These constraints have changed our guideline to construct the high density part of EoS.

The constraints on the high density EoS, however, by themselves do not powerfully constrain theoretical scenarios on the properties of matter; we need to discuss how they are connected to the low density EoS. For the soft-to-stiff combination, the radius tends to be small while the maximum mass is large; the resulting stars are very compact and baryon density tends to be high. This type of EoS is hard to accommodate the strong first order phase transition in the domain $2$-$5n_0$; starting with soft EoS,  the first order phase transition and the associated softening easily makes NS unstable to the gravitational collapse. In contrast, for the stiff-to-stiff combination, the high density EoS can be stiff even after the first order transition so that it can satisfy the $2M_\odot$ constraint for variety of high density EoS. 
Starting with very stiff low density EoS, the overall radius tends to be large, and if there are first order transitions, the radius in the $M$-$R$ curve shrinks rapidly. The question of the existence of the strong first order phase transitions is directly related to the estimate of the NS radius or the low density EoS.

The estimate of the NS radii has been more uncertain than of the NS mass. Before the detection of the NS merger, there were three sources of information: 
(i) the spectroscopic analyses of the X-rays from the neutron star surface constrains on the radii of typical NS with $\sim 1.4M_\odot$ \cite{timing}. The current trend is converging toward the estimate $R_{1.4}=10$-$13$ km. The major problem is the systematic uncertainties;
(ii) the heavy ion collisions at low energy with a few GeV have the sensitivity to the EoS  \cite{Li:2008gp,Danielewicz:2002pu}. The collective flow and particle production depend on how much two nuclei get compressed. They indicate soft EoS around $\sim 2n_0$ which leads to $R_{1.4}=10$-$13$ km. This discussion also includes the systematic uncertainties;
(iii) the laboratory experiments for nuclei measure nuclear EoS around $n_0$ and its density derivatives. There have been significant progress in the estimate for $\sim n_0$, but we need extrapolations to the domain relevant for NS, introducing significant errors for $\gtrsim 2n_0$.

%%%%%%%%%
\begin{figure}
\begin{center}
\vspace{-1.0cm}
\hspace{-2.5cm}
\includegraphics[width=8.0cm,bb=0 0 600 600, angle =0]{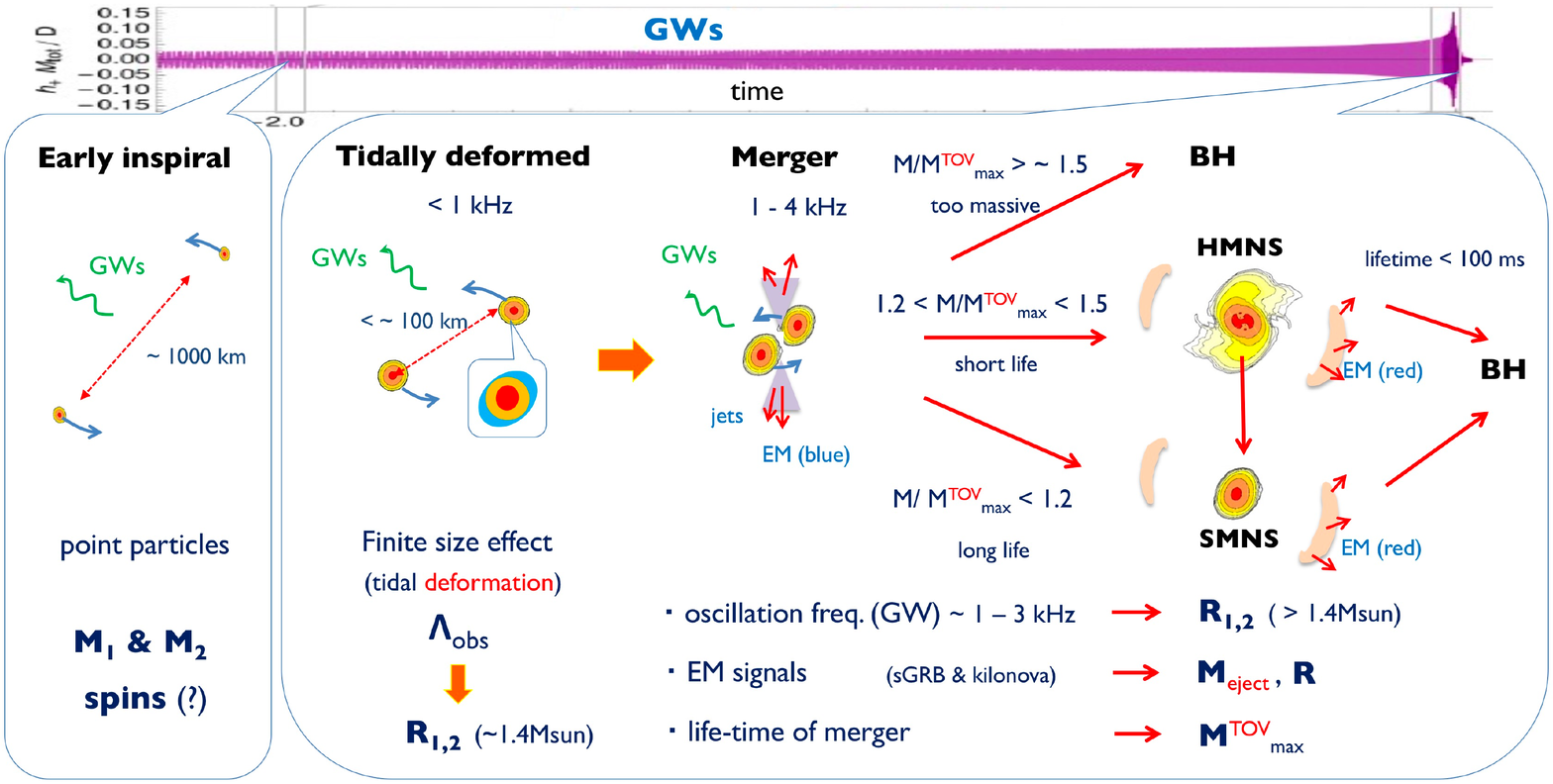}
\vspace{-1.cm}
\caption{
\footnotesize{ Time evolution of NS-NS mergers.
}
  }
  \vspace{-0.5cm}
  \end{center}
 \label{fig:GWs}
\end{figure}
%%%%%%%%%

These constraints for the $M$-$R$ curves have been considerably improved by the discovery of the NS merger event GW170817, found in Aug.17, 2017. The total mass is $M_{ GW170817} =2.73$-$2.78 M_\odot$ and it is plausible to have the mass ratio $M_1/M_2$ of $0.7$-$1.0$ with which $M_{1,2} \simeq 1.3$-$1.4M_\odot$. It is a very rich source of information, as it contains several different evolution stages, see Fig.\ref{fig:GWs}: 

 In its early stage, two NSs are widely separated and they can be regarded as point particles. This binary motion as a whole contains the oscillation of the quadrupole component which is the source for the GW emissions. 

This point particle regime continues for a long time and eventually two NSs come close together;
at this stage the finite size effects set in. A NS is deformed by the gravitation fields due to the other, and such deformation (tidal deformation) in turn induces net attractive forces, accelerating the merger process, and then increasing the frequency of GWs. The tidal deformability is very sensitive to the radius of NS and thus strongly constrains the NS radii around $\sim 1.4M_\odot$.

 When the two NS merge, the system enters the highly nonlinear regime described by the descriptions based on the general relativity, magnetohydrodynamics, and neutrino transport, together with EoS for general charge fractions and temperatures. This stage contains pretty rich information, but unfortunately right now the GWs from this high frequency domain ($\gtrsim$ 1kHz) have not been detected with the detector sensitivity achieved in O2. Nevertheless there are already several valuable information for this dynamical stage.
 
%%%%%%%%%
\begin{figure}
\begin{center}
\vspace{-1.0cm}
\hspace{-2.5cm}
\includegraphics[width=7.5cm,bb=0 0 600 600, angle =0]{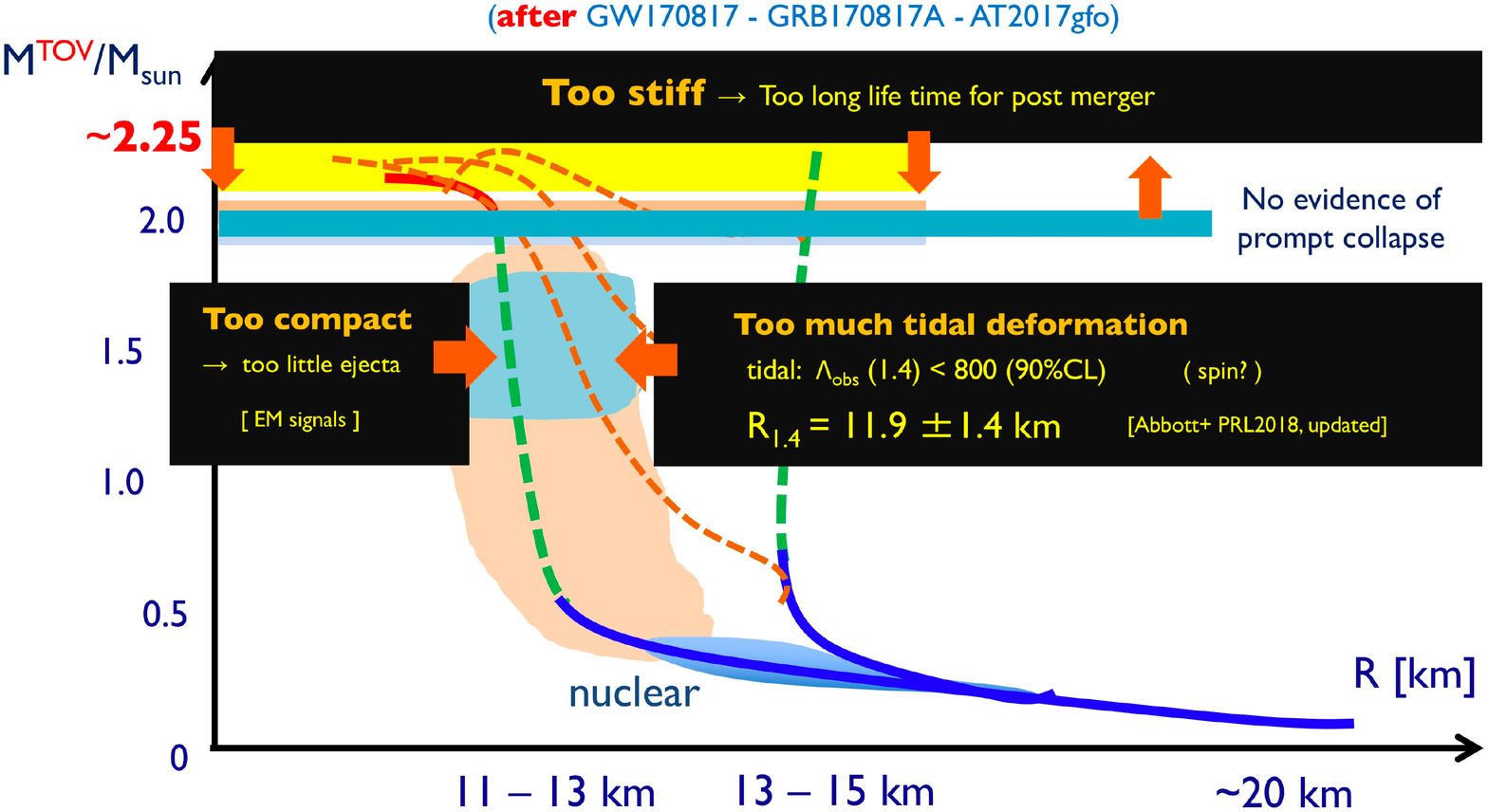}
\vspace{-1.cm}
\caption{
\footnotesize{ Constraints from the GW170817 event. The black regions are excluded.
}
  }
  \vspace{-0.5cm}
  \end{center}
 \label{fig:GWconstraints}
\end{figure}
%%%%%%%%%

The lifetime of the merged object has been inferred from electromagnetic counterparts. The lifetime can be used to place the lower as well as upper bound on the $M_{\rm max}$. If the merged objects are too massive ($ M/M_{{\rm max}} \gtrsim 1.5$) compared to the maximal mass of a non-rotating NS, the object promptly collapses to the BH. If not too massive ($1.5 \gtrsim M/M_{{\rm max}} \gtrsim 1.2$), the merger forms a differentially rotating metastable object, a hypermassive NS, which survives for a while and then collapses. If the mass is relatively light  ($1.2 \gtrsim M/M_{{\rm max}} \gtrsim 1$), then the star survives for a long time as a uniformly rotating quasistable object, supramassive NS. 
The amount of the mass ejecta and electromagnetic counterparts seems to favor the hypermassive NS scenario for the GW170817  \cite{Margalit:2017dij,Rezzolla:2017aly,Ruiz:2017due,Shibata:2017xdx} (see also \cite{Yu:2017syg} for an alternative scenario). 
The inequality $1.5 \gtrsim M_{GW170817}/M_{{\rm max}} \gtrsim 1.2$ leads to $1.82 \lesssim M_{\rm max}/M_\odot  \lesssim 2.32$.
The ejecta of $\sim 10^{-1}$-$10^{-2} M_\odot$ also indicates that the NS should not be too compact; otherwise the ejected materials would be largely swallowed when the BH is formed. This places the lowerbound on the NS radius around $R \gtrsim 10$-$11$ km. If GWs in the high frequency domain is directly measured, the estimate of the lifetime becomes more accurate and improves the confidence level of all these estimates. In addition, the high frequency oscillation also is sensitive to the NS radii for $M\gtrsim 1.4M_\odot$; we can obtain the information of $M$-$R$ curves near $M_{\rm max}$. This will be useful to check whether the $M$-$R$ curve has a kink for $M\gtrsim 1.4M_\odot$. It can be a signature of the 1st order phase transitions between $\sim 2n_0$ and $\sim 5n_0$.

Shown in Fig. \ref{fig:GWconstraints} is the overall summary of the constraints obtained from the GW170817.
For the maximum mass based on the hypermassive NS picture; $M_{\rm max} \lesssim 2.17 M_\odot$ (Margalit and Metzger \cite{Margalit:2017dij}); $\lesssim 2.16^{+0.17}_{-0.15} M_\odot$ (Rezzola et al. \cite{Rezzolla:2017aly}); $\lesssim 2.16 M_\odot$ (Ruiz et al. \cite{Ruiz:2017due}); $2.15$-$2.25 M_\odot$ (Shibata et al. \cite{Shibata:2017xdx}). 
Meanwhile, there is a possibility that the GW170817 has the merger as the long-lived supramassive star. In this case we obtain totally different conclusion with $M_{\rm max}/M_\odot  \gtrsim 2.32$ (Yu et al. \cite{Yu:2017syg}). As for the NS radii, the lower bound was given by the condition for sufficient amount of ejecta consistent with the electromagnetic observations; $R_{1.6} \ge 10.68^{+0.15}_{-0.04}$ km (Bauswein et al. \cite{Bauswein:2017vtn}); $R_{1.4} \gtrsim 11.0$-$11.5$ km (Radice et al. \cite{Radice:2017lry}) gave the constraint $\tilde{\Lambda}_{1.4} \gtrsim 400 $. Meanwhile the upper bound is derived from the tidally deformed phase. The upper bound of $\tilde{\Lambda}$ was directly estimated by the LIGO from the analyses of gravitational waves as $\Lambda_{1.4} \lesssim 800$ (CL90\%), and it is translated to the radius constraint $R_{1.4} \lesssim 13.6$ km. 
Later the estimate was revised to $R_{1.4} = 11.9\pm 1.4$ km (CL90\%) \cite{Abbott:2018exr}. Useful summary in terms of $M$-$R$ curves can be found in \cite{Bauswein:2017vtn,Annala:2017llu}.

Integrating these information, below it seems reasonable to assume the QCD EoS to be soft-to-stiff type. Below we assume it and discuss the outcome of this assumption.

%%%%%%%%%%%%%%%%%%%%%%%%%%%%%%%%%%%%%%%%%%%%%%%%%%%%%%
\section{Soft-to-stiff equations of state and quark-hadron continuity }
%%%%%%%%%%%%%%%%%%%%%%%%%%%%%%%%%%%%%%%%%%%%%%%%%%%%%%

%%%%%%%%%
\begin{figure}[tb]
%\begin{center}
\vspace{-1.0cm}
\centering
\includegraphics[scale=0.2]{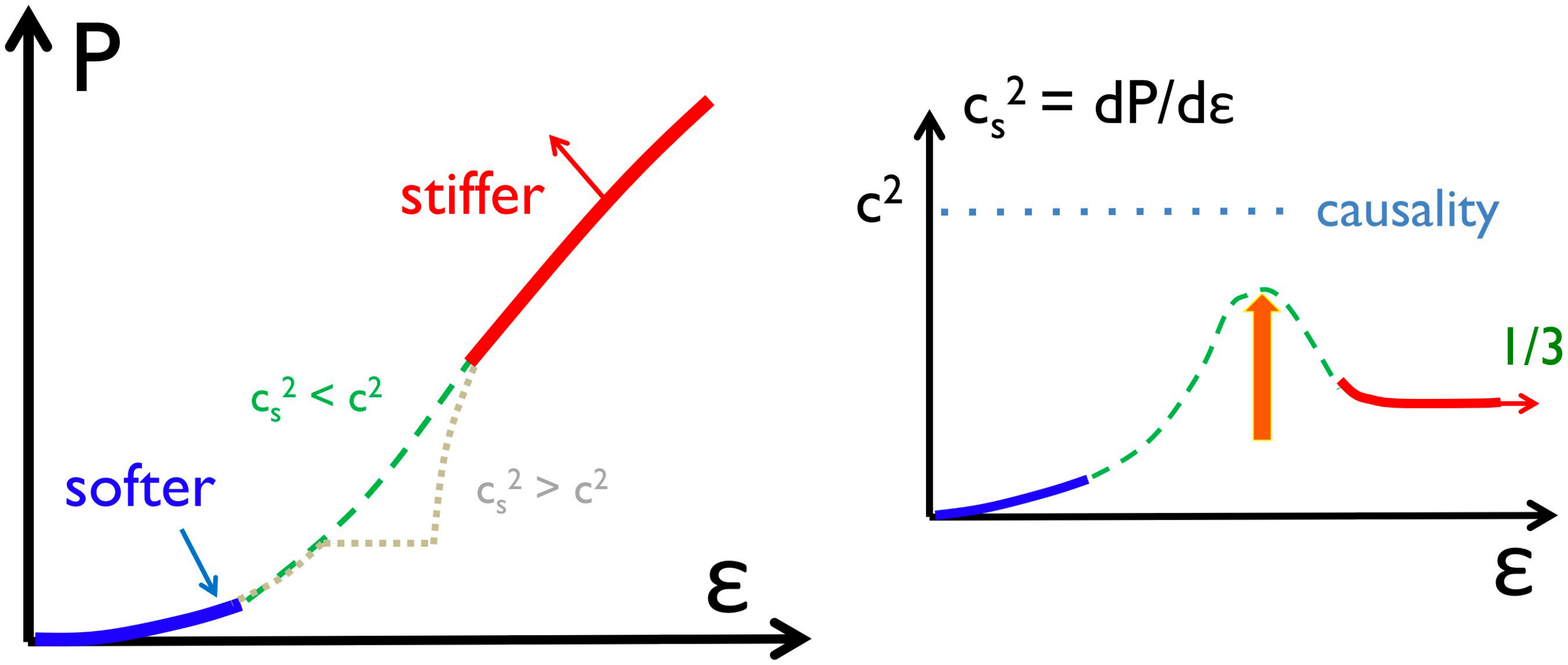}
\vspace{-1.cm}
\caption{
The pressure vs energy density and speed of sound with or without the first order phase transitions.
  }
  \vspace{-0.cm}
%  \end{center}
   \label{fig:cs2}
\end{figure}
%%%%%%%%%
%%%%%%%%%%%%%%%%%%%%%%%%%%%%%%%%%%%
%%%%%%%% Speed of sound

The soft-to-stiff EoS has a rather peculiar structure (Fig.\ref{fig:cs2}). In order to connect soft EoS at low density to stiff ones at high density, the pressure grows rapidly as a function of energy density. The speed of sound becomes large for the domain $2$-$5n_0$. But the causality condition requires that the speed of sound must be smaller than the light velocity. This constrains the structure of $P(\varepsilon)$ curves. 
The soft-to-stiff EoS demands the appearance of peaks in $c_s^2$ exceeding the conformal limit $1/3$, provided that the EoS eventually approaches the pQCD result with $c_s^2 \simeq 1/3$ at high density. If there are first order phase transitions, the associated softening demand the other domain to have even stronger peaks, but stronger peaks tend to violate the causality bound \cite{Baym:2017whm,Kojo:2015fua}. For more systematic arguments we refer to \cite{Bedaque:2014sqa,Alford:2013aca,Tews:2018kmu}.
 
This situation leads us to the picture of quark-hadron continuity; there is no sharp thermodynamic phase transitions between hadronic and quark phases. 

Such continuity picture was originally discussed in the context of symmetries; transitions from the superfluid hadronic phase to the color-flavor-locked superconducting phase \cite{Schafer:1998ef}. 
In fact the quantum numbers carried by hadronic degrees of freedom find their counterparts in quark matter. This scenario was revisited with questions concerning the dynamics \cite{Hatsuda:2006ps,Zhang:2008wx}, where the interplay between the chiral and diquark condensates plays the key role. These studies are based on theoretical considerations and model calculations. In contrast we have reached the continuity picture in attempts to be consistent with the NS constraints.

Our way of using the terminology ``continuity" is looser than the previous theoretical studies; right now the strict connection between order parameters in hadronic and quark matter is not our primary concern, unless these order parameters have significant impacts on EoS. For instance the appearance of gaps of nuclear scale $1$-$10$ MeV in superfluid phases (for a recent review \cite{Sedrakian:2018ydt}) is, for the moment, not crucial ingredient in our arguments.  They will become more critical when we proceed to the discussion of the neutron star cooling (for review, e.g., \cite{Yakovlev:2004iq}).

While we are relatively loose in discussions for order parameters, we do care how the dynamics relevant for the hadron and nuclear physics evolve from low to high density. We will come back to this point when we introduce a schematic quark model.

%%%%%%%%%%%%%%%%%
\section{The 3-window modeling}
%%%%%%%%%%%%%%%%%

%%%%%%%%%
\begin{figure}[tb]
%\begin{center}
\vspace{-0.5cm}
\centering
\includegraphics[scale=0.2]{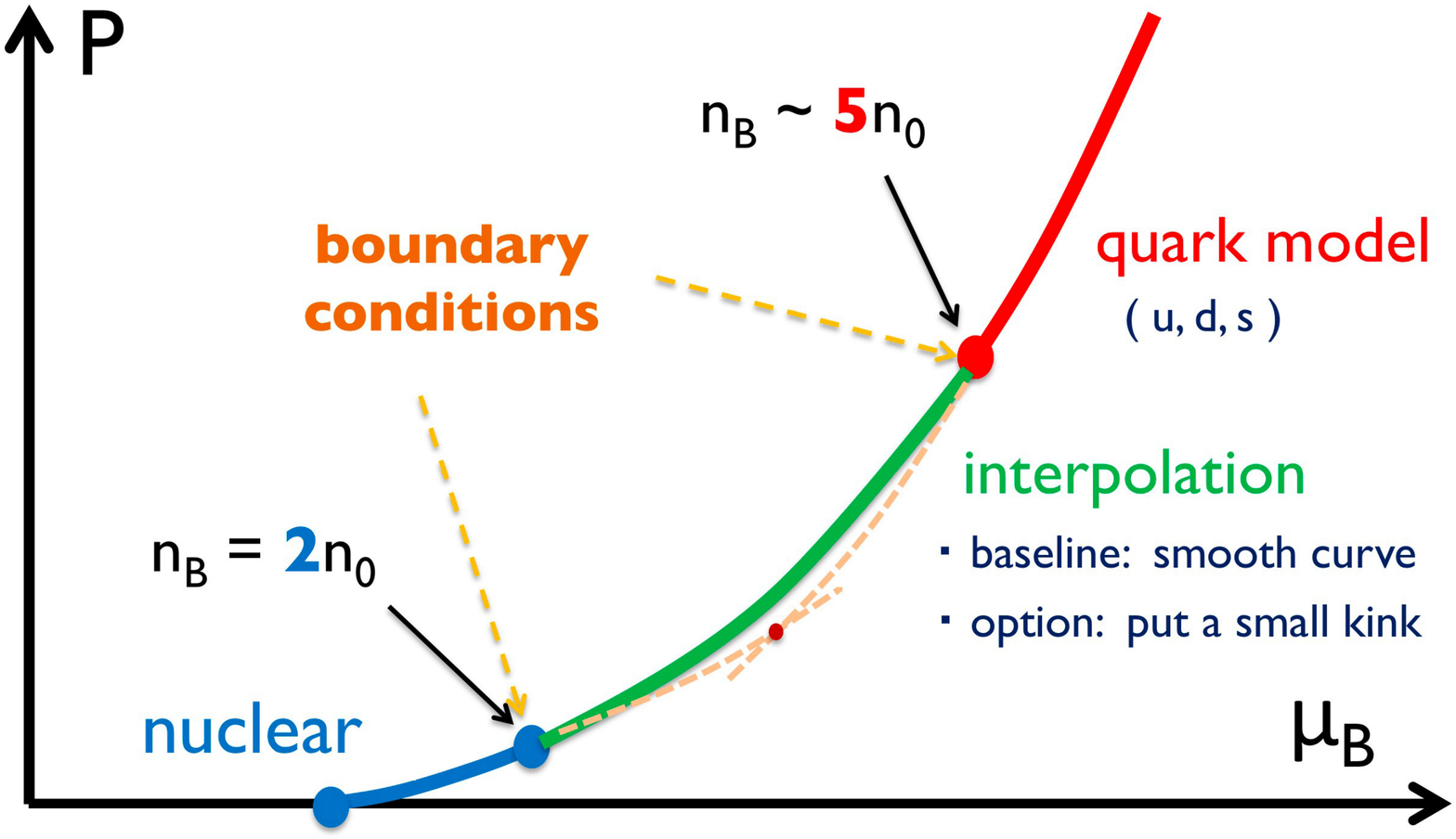}
\vspace{-0.5cm}
\caption{
The 3-window construction of the unified EoS.
  }
  \vspace{-0.cm}
%  \end{center}
   \label{fig:cs2}
\end{figure}
%%%%%%%%%

After getting hints for the quark-hadron continuity picture, now we implement the idea using the 3-window picture.
We use this 3-window model to extract the physical insights out of the EoS supposed from the NS observations.
We use a nuclear EoS to $n_B \sim 2n_0$ beyond which stop using it. At high density we use a quark EoS but stop using it below $\sim 5n_0$ because of confining effects not controllable.
Taking the nuclear and quark EoS as the boundary conditions at $2n_0$ and $ 5n_0$, we interpolate these EoS with polynomials \cite{Kojo:2014rca}
\beq
P(\mu_B) = \sum_{n=0}^5 c_n \mu_B^n \,.
\eeq
In this expression we have six coefficients, as we will demand the matching conditions up to the second derivatives of $P$ with respect to $\mu_B$. These boundary conditions uniquely fix $c_n$'s.
\footnote{Clearly the form chosen is just one choice of interpolating functions. If we wish, we can use pressure curves including small kinks to describe the first order phase transitions, but the effects of kinks should be small according to the constraints discussed in the previous section. Therefore we take smooth curves to be our baseline. } 

There are requirements for the interpolating curves. They must be such that (i) $P(\mu_B)$ is convex and has the positive curvature everywhere; (ii) $P(\mu_B)$ must lead to the causal speed of sound $c_s \le 1$; in turn the growth of $P(\mu_B)$ cannot be too slow for increasing $\mu_B$. The pressure curves which do not satisfy these constraints are unphysical. 

At this point it is important to emphasize the difference between the three-window modeling and the hybrid construction with first order hadron-quark phase transitions. The latter is based on the assumption that the hadronic and quark matter are distinct. The location of the phase transition point is determined by direct comparison of pure hadronic and pure quark matter EoS. Tacit assumption of this procedure is that both EoS are reliable. But as we have estimated, the domain of $2$-$5n_0$, where we typically locate the first order phase transitions, does not allow reliable use of pure nucleonic nor pure quark matter descriptions. But perhaps more important feature is that, by extrapolating the nuclear EoS beyond the applicability, one would artificially reject physically sensible type of quark EoS when we try to describe the hadron-quark transitions. In the opposite case, one extrapolate quark EoS to too low density and would find a problem with the nuclear EoS, then disregard the entire domain of such quark EoS; but this might not be a problem of the quark EoS, but simply be of the wrong extrapolation. In either case EoS based on extrapolated curves has danger to introduce unphysical constraints on the properties of quark matter. 

To avoid such biases, it is safer to use the 3-window construction in which nuclear and quark EoS are used only within the domain of reliability.
In this approach, we avoid the biases, and in turn can consider a class of quark EoS which have been rejected previously. In particular they can accommodate stiff quark EoS even when we start with soft hadronic EoS at low density. 

Specifically we use the Togashi EoS for the nuclear domain at $n_B \lesssim 2n_0$ \cite{Togashi:2017mjp}. The calculations are based on a Hamiltonian including two-body interactions extracted by fitting two-nucleon experimental scattering data,  as well as more empirical three-body interactions fit to light nuclei. With these inputs Togashi et al. solved the many-body problem by choosing a variational wave function with parameters determined by minimizing the total energy. The results of the calculations are consistent with other microscopic computations based on the chiral effective theories. 

Interpolating the Togashi EoS and quark EoS described below, we construct the unified EoS. We call it the Quark-Hadron-Crossover EoS (QHC19).

Here it is instructive to compare our interpolation with another one in which chiral effective theory based EoS of $\lesssim 1.1 n_0$ and pQCD one of $\gtrsim 50 n_0$ are interpolated \cite{Annala:2017llu}. 
The purposes of these approaches are different. The latter aims at the prediction of EoS between $\sim 1.1n_0$ and $\sim 50n_0$. In contrast, in our interpolation between $\sim 2n_0$ and $\sim 5n_0$, our goal is not to predict the EoS, but delineate the properties of QCD matter that are consistent with the NS observations. For this purpose we use a schematic quark model whose effective interactions are motivated from the hadron spectroscopy, and analyze the impact of such interactions on the EoS. This procedure provides us with the insights into the effective interactions and the emerging picture of QCD matter.

%%%%%%%%%%%%%%%%%
\section{Delineating the properties of matter through a schematic quark model}
%%%%%%%%%%%%%%%%%

%%%%%%%%%
\begin{figure}[tb]
%\begin{center}
\vspace{-1.0cm}
\centering
\includegraphics[scale=0.23]{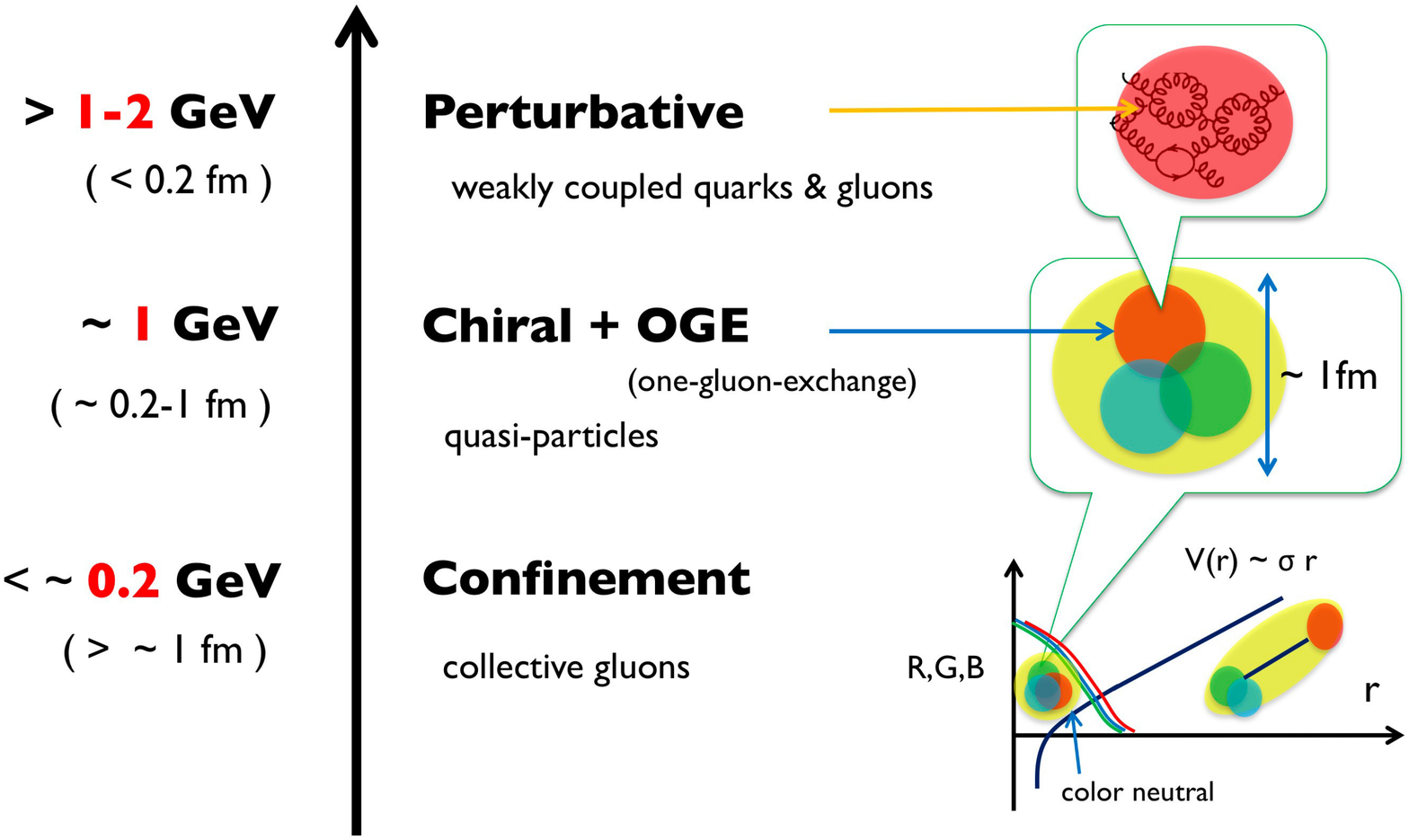}
\vspace{-0.5cm}
\caption{
The 3-window picture of a single hadron.
  }
  \vspace{-0.cm}
%  \end{center}
   \label{fig:cs2}
\end{figure}
%%%%%%%%%

We first guess the relevant physics at $n_B \gtrsim 5n_0$. Here we consider another ``3-window" description for a single hadron. The picture is inspired from the arguments by Manohar and Georgi \cite{Manohar:1983md} and later by Weinberg \cite{Weinberg:2010bq}. The 3-window consists of the physics of (i) the confinement for the momentum scale of $\lesssim 0.2$ GeV (or distance $\gtrsim 1$ fm); (ii) the constituent quark dynamics  for 0.2-1 GeV (0.2-1 fm); (iii) the partonic dynamics for $ \gtrsim 1$ GeV ($\lesssim 0.2$ fm).

Now the most relevant for our studies of the NS domain is supposed to be the regime (ii). The relevant ingredient is the chiral symmetry breaking and one-gluon-exchange between quarks. The gluon exchange with momenta  of 0.2-1 GeV is replaced with contact interactions for which we introduce the cutoff scale of $\sim 1$ GeV. 
Now our effective Hamiltonian is ($\mu_q =\mu_B/3$) \cite{Baym:2017whm}
\begin{eqnarray}
\calH  
= \bar{q} (\rmi \gamma_0 \vec{\gamma}\cdot \vec{\partial} + m -\mu_q \gamma_0)q 
- G_s \sum^8_{i=0} \left[ (\overline{q} \tau_i q)^2 + (\bar{q} \rmi \gamma_5 \tau_i q)^2 \right] 
+ 8 K ( \det\,\!\!_{f} \bar{q}_R q_L + \mbox{h.c.}) \nonumber \\
%+ \calH_{ conf }^{  3q\rightarrow B }  
- H \!\sum_{A,A^\prime = 2,5,7} \!
 \left(\bar{q} \rmi \gamma_5 \tau_A \lambda_{A^\prime} C \bar{q}^T \right) \left(q^T C \rmi \gamma_5 \tau_A \lambda_{A^\prime} q \right) + g_V (\overline{q} \gamma^\mu q)^2
 \,.
 \label{eq:H}
\end{eqnarray}
The first line is the standard Nambu--Jona-Lasinio (NJL) model with $u,d,s$- quarks and responsible for the chiral symmetry breaking. We use the Hatsuda-Kunihiro parameter set \cite{Hatsuda:1994pi} which leads to the dynamically generated quark masses of $M_{u,d} \simeq 336 $ MeV and $M_s \simeq 528$ MeV as functions of the current quark mass matrix $m$, the scalar coupling $G_s$, and the coefficient of the Kobayashi-Maskawa-'t Hooft vertex, $K$. The meson phenomenology has been successfully described within the Hamiltonian in the first line of Eq.(\ref{eq:H}). Meanwhile the terms in the second line are more important for baryons. The first term includes the color magnetic interaction for color-flavor-spin antisymmetric S-wave interaction which is attractive. The last term is a vector repulsive interaction which is originally inspired from the $\omega$-meson exchange in nuclear physics. From more modern point view, the short distance repulsion has its microscopic foundation in color-magnetic interactions as predicted by \cite{Oka:1980ax} and supported by the lattice studies \cite{Park:2019bsz}.

While the form of the Hamiltonian is obtained by extrapolating the description of hadron and nuclear physics, in principle the range of parameters $(G_s, K, g_V, H)$ at $n_B \ge 5n_0$ can be considerably different from those used in hadron physics due to e.g. medium screening effects. In strongly correlated region the estimate of medium modifications is difficult; for instance screening masses in 2-color QCD, measured in lattice QCD \cite{Hajizadeh:2017ewa}, are qualitatively different from the perturbative behaviors \cite{Kojo:2014vja}. For 3-color QCD, no quantitative estimates are available, so here we use the NS constraints to examine the range of these parameters to delineate the properties of QCD matter at $n_B \ge 5n_0$. Below we vary ($g_V, H)$, while assume that $(G_s, K)$ do not change from the vacuum values appreciably; this assumption will be favored posteriori. More elaborated treatment is to explicitly treat the medium running coupling $g_V(\mu_B)$, as demonstrated in Ref.\cite{Fukushima:2015bda}. The attempt to estimate $g_V$ and $H$ from the high density regime can be found in \cite{Song:2019qoh}; the estimates show remarkable agreement with the estimate based on the NS physics (see below).

Our Hamiltonian for quarks, together with the contributions from leptons, is solved within the mean field 
approximation\footnote{ In order to calculate the crossover region from hadronic to quark matter, one must use a unified framework which can accommodate both quarks and hadrons. One of theoretical problems for such framework is the treatment of the zero point energy of quarks and hadrons which would separately diverge. 
Actually we can show that the sum of those divergent contributions is zero, provided that certain conditions are satisfied \cite{Kojo:2018usw}. But its practical implementation demands considerable works and has not been completed. Most typically the zero point energy is dropped off by hand (no sea approximation).
}. 
As usual we use the Lagrange multipliers to impose the neutrality conditions for electric and color charges as well as the $\beta$-equilibrium condition \cite{Baym:2017whm}. In the mean field treatments we find that the chiral and diquark condensates coexist at $n_B \ge 5n_0$, and the diquark pairing appears to be the color-flavor-locked (CFL) type.

%%%%%%%%%
\begin{figure}[tb]
%\begin{center}
\vspace{-0.7cm}
\centering
\includegraphics[scale=0.5]{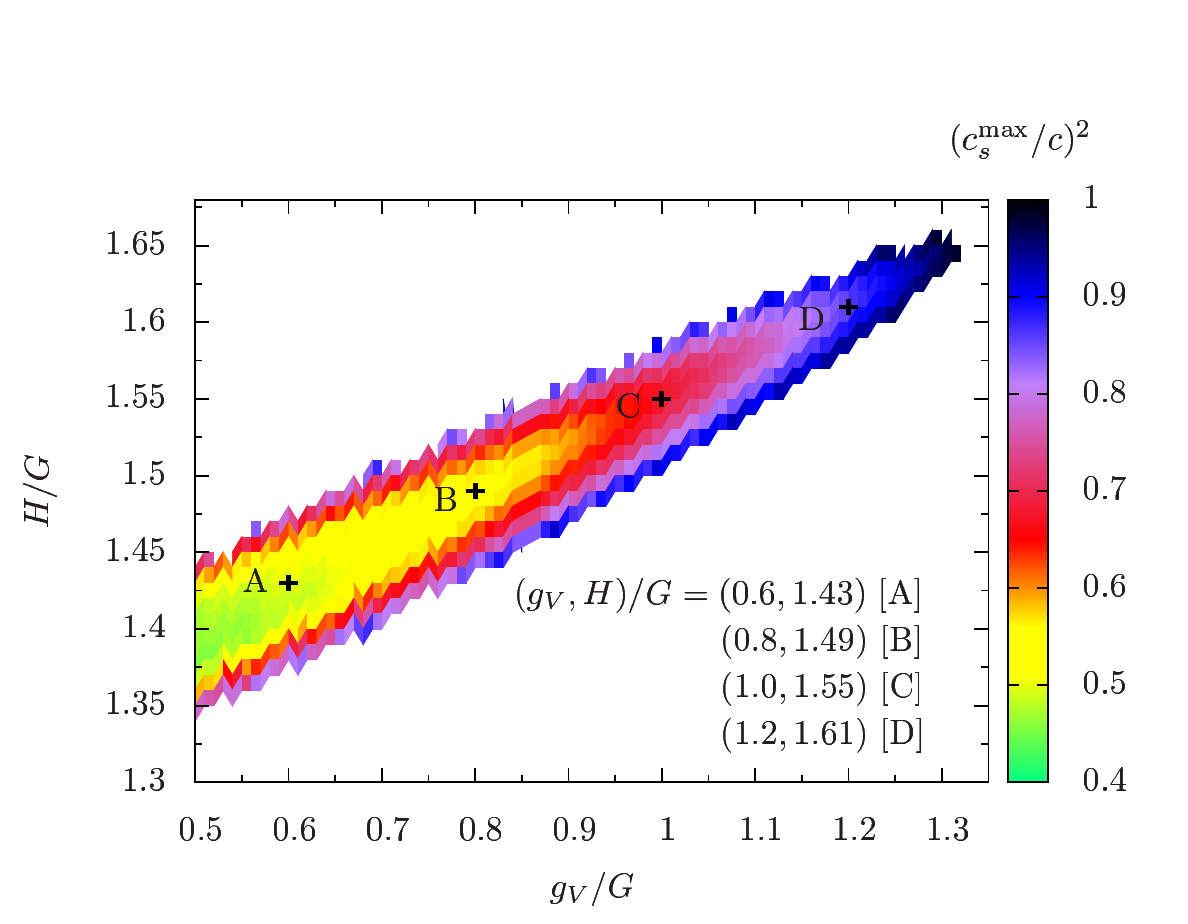}
\hspace{-0.cm}
\includegraphics[scale=0.5]{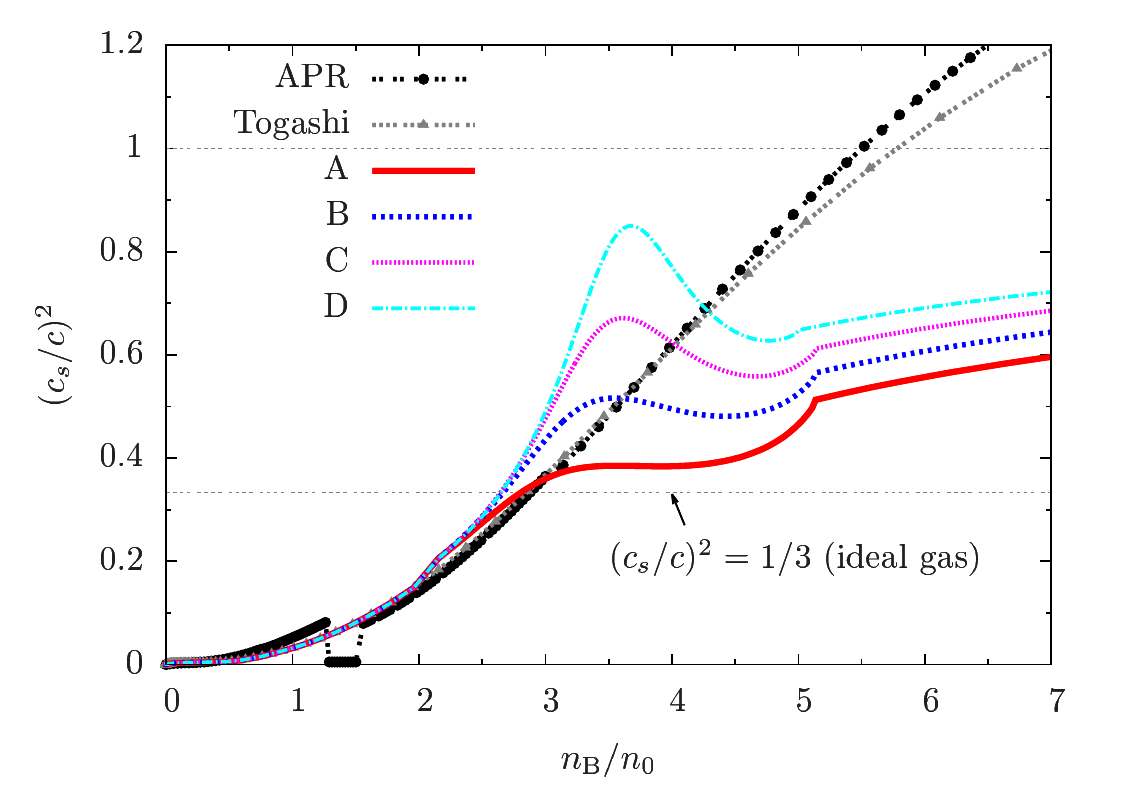}
%\vspace{-0.3cm}
\caption{ (Left) The maximum value of speed of sound square, $c_s^2$, for a given $(g_V,H)$, found in the interval 2-5$n_0$. The domain left blank is unphysical. The samples A-D are chosen for the right panel which includes also the speed of sound for the APR and Togashi EoS.
  }
  \vspace{-0.cm}
%  \end{center}
   \label{fig:cs2_gv_H}
\end{figure}
%%%%%%%%%

First we specify the range where the speed of sound is physical. Shown in the left panel in Fig.\ref{fig:cs2_gv_H} is the maximum of the sound velocity in the interval 2-5$n_0$. The domain left blank is unphysical. There is a strong correlation between the range of $g_V$ and $H$. As samples of $(g_V,H)/G_s$, we take (0.6, 1.43) (set A); (0.8, 1.49) (set B); (1.0, 1.55) (set C); and (1.2, 1.61) (set D). The speed of sound square as functions of $n_B$ for sets A-D are shown in the right panel in Fig.\ref{fig:cs2_gv_H}. Our QHC19 for the sets A-D have peaks or bumps for 2-5$n_0$. The direct approach to the interpolated domain with the peak structure has been attempted in \cite{McLerran:2018hbz,Jeong:2019lhv} based on the quarkyonic matter picture.

\begin{figure}[tb]
%\begin{center}
\vspace{-0.5cm}
\centering
\includegraphics[scale=0.6]{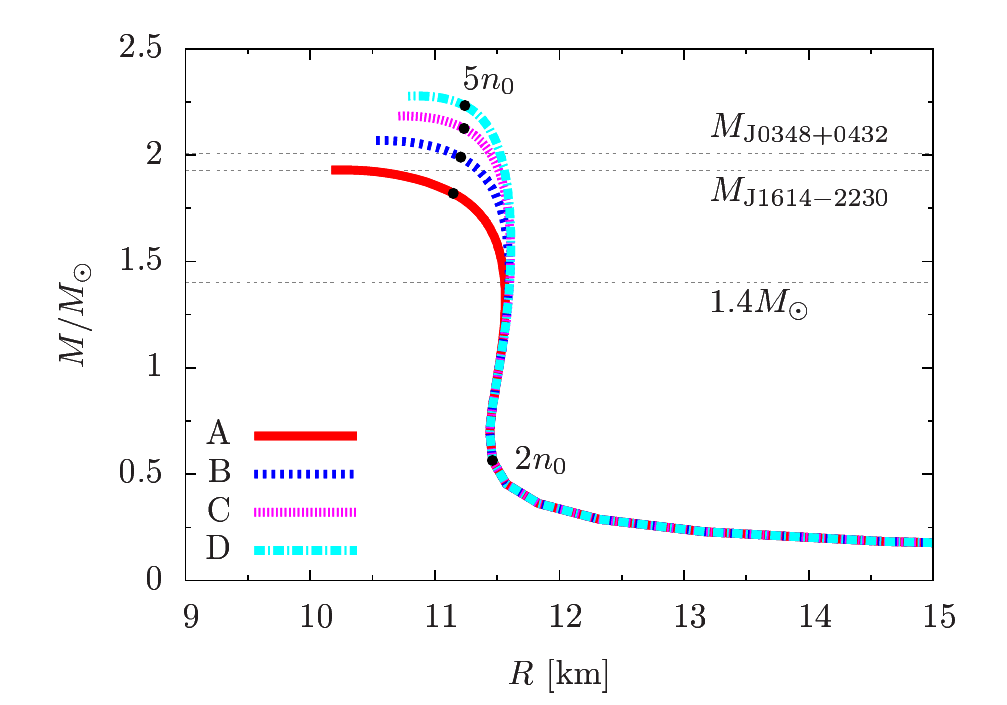}
\hspace{-0.cm}
\includegraphics[scale=0.5]{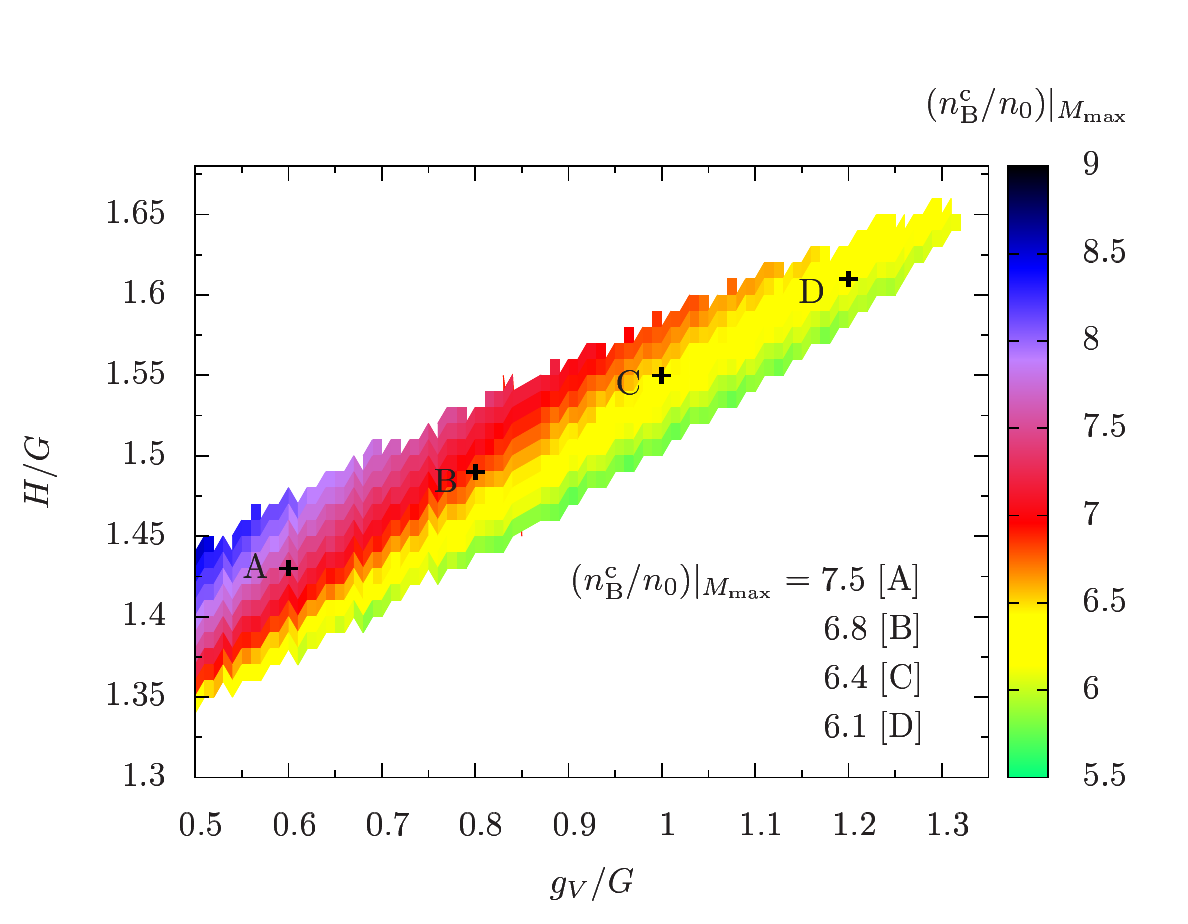}
%\vspace{-0.3cm}
\caption{ The $M$-$R$ relations (left) and the core baryon density (right) of NSs for a given $(g_V,H)$. 
  }
  \vspace{-0.2cm}
%  \end{center}
   \label{fig:Mmax}
\end{figure}

Shown in Fig.\ref{fig:Mmax} is the $M$-$R$ relations for the set A-D. The radius $R_{1.4}$ of a NS is $\simeq 11.6$ km, mainly determined by the Togashi EoS. The quark model parameters varied for the entire allowed range affect $R_{1.4}$ by $\simeq 0.5$ km at most. Meanwhile the maximum mass is sensitive to the quark model parameters.
It shows that $g_V/G_s$ should be larger than 0.6-0.7, otherwise the maximum mass does not exceed $2M_\odot$. Correspondingly $H/G_s$ must be bigger than $\sim 1.35$. We found that such choice of $H$ leads to the diquark gap of $\sim$ 200-250 MeV at $n_B \sim 5n_0$. Within our approach the largest maximum mass is found to be $\simeq 2.35M_\odot$ where the domain of $H$ shrinks to zero.

%%%%%%%%%
%%%%%%%%%

Shown in the right panel of Fig.\ref{fig:Mmax} is the core baryon density, $n_B^c$. For a given $g_V$, a larger $H$ leads to the larger $n_B$. This is because the diquark pairing favors the larger Fermi surface to create pairs as much as possible to reduce the energy of the system. Meanwhile larger $g_V$ for a given $H$ reduces $n_B^c$ to avoid the energy cost. In the entire domain of $(g_V,H)$ for $g_V> 0.5G$ the core baryon density is larger than $5n_0$, while the upperbound comes from the $2M_\odot$ constraints that set the core density less than $\simeq 8n_0$. 

After constraining the range of effective interactions we now can discuss more microscopic aspects of the matter. We found the following trends: (i) for $2M_\odot$ NS, the core density can reach $n_B \gtrsim 5n_0$ or $\mu_B \gtrsim 1.5$ GeV. At such high density the appearance of the strangeness seems unavoidable and we find almost equal population of up-, down-, and strange-quarks, $n_u \simeq n_d \simeq n_s$; (ii) the pairing gap for the CFL phase appears to be $\sim 200$ MeV, although this estimate is model dependent and the further examination is necessary; (iii) the repulsive density interactions temper the growth of the baryon density significantly. This in turn temper the restoration of the chiral symmetry. At $\simeq 5n_0$, our model predicts the effective quark mass $M_{u,d} \simeq 50$ MeV and $M_s \simeq 300$ MeV; the substantial chiral symmetry breaking remains at the NS cores.

%%%%%%%%%%%%%%%%%
\section{Summary}
%%%%%%%%%%%%%%%%%

Recent NS observations provide us with clues to understand the cold dense matter in QCD. Last ten years we have witnessed the dramatic changes in the $M$-$R$ constraints. In particular the GW170817 event put the lower and upper bounds on the maximum mass as well as the radius. The impact of this single event is already remarkable, but more events will come in next ten years and enable us to perform statistical analyses to improve our confidence of the constraints. These constraints then must be translated into the language of microphysics. The insights on the microphysics will allow us to improve EoS and to predict other quantities such as transport coefficients which are relevant for dynamics of NS or cooling of NS. The latter is useful to disentangle theoretical scenarios on the QCD phase structure. For the exploration of the microphysics, it is very valuable to study the QCD like theories which share some aspects with QCD and can be computed in lattice simulations.
\\

The author thanks the organizers of the conference for this very enjoyable meeting. He also thanks G. Baym, K. Fukushima, S. Furusawa, T. Hatsuda, D. Hou, J. Okafor, P. Powell, Y. Song, D. Suenaga, T. Takatsuka, H. Togashi, for collaboration. 
This work is supported by the NSFC grant 11650110435 and 11875144.

%%%%%%%%%%%%%%%%%%%%%

\end{document}